# A New Theoretical Interpretation of Measurement Error and Its Uncertainty

Huisheng Shi [1], Xiaoming Ye [2*], Cheng Xing [2] and Shijun Ding [2]

[1] AVIC Changcheng Institute of Metrology & Measurement, Beijing, 100095, P.R. China.
[2] School of Geodesy and Geomatics, Wuhan University, Wuhan, 430079, P.R. China.

Correspondence: Xiaoming Ye; xmye@sgg.whu.edu.cn

## Abstract

The traditional measurement theory interprets the variance as the dispersion of a measured value, which is actually contrary to a general mathematical concept that the variance of a constant is 0. This paper will fully demonstrate that the variance in measurement theory is actually the evaluation of probability interval of an error instead of the dispersion of a measured value, point out the key point of mistake in the traditional interpretation, and fully interpret a series of changes in conceptual logic and processing method brought about by this new concept.

## 1. Introduction

Human scientific research begins with the measurement of various physical quantities, and measurement is the basis of modern scientific research. Especially for the research of artificial intelligence, it is an essential process to obtain natural information and evaluate its authenticity, which itself is the research area of measurement. Therefore, a rigorous measurement theory should be an important part of the modern scientific theory system. This paper points out the main problems existing in the traditional measurement theory and gives the correct interpretation method for the measurement theory.

In traditional measurement theory, the measured value (or observed value) is considered as a random variable and variance is interpreted as the dispersion of the measured value, both the precision and uncertainty are defined as the dispersion concept of measured value (or observed value) [1, 2, 3, 4, 5], so that people can hardly make clear the conceptual difference between them.

However, in any measurement, both the measured value and every observed value are numerical values. According to probability theory, the variance of a numerical value (constant) is zero. So, how does a numerical value show dispersion? Next, we illustrate the contradictory expression of “variance” in traditional theory.

For example, the measured value of Mount Everest elevation in 2005 is $x = 8844.43\text{m}$, and its precision is $\sigma(x) = 0.21\text{m}$. But in fact, this mathematical expression gives a wrong equation $\sigma(8844.43\text{m}) = 0.21\text{m}$, which violates basic mathematical concept, because the equation $x = 8844.43\text{m}$ inevitably leads to the equation $\sigma(x) = \sigma(8844.43\ )$, and according to the concept of $\sigma^2(C) = 0$ in probability theory, there must be $\sigma(8844.43) = 0$. Although many other measured values $x_1$, $x_2$, ..., can be obtained by repeatedly measuring the height of Mount Everest, and there can be $x \neq x_1 \neq x_2 \neq \cdots$, the equations $x = 8844.43\text{m}$ and $\sigma(x) = \sigma(8844.43)$ still exist. Therefore, the equation $\sigma(x) = \sigma(8844.43) = 0.21$ can never be consistent with mathematical concepts.

It can be seen that in traditional theories, the definition of precision and uncertainty is the dispersion of all possible measured values, but their mathematical expression is the dispersion of a single measured value (a single numerical value). This approach is actually a stealth change of concept, and will inevitably lead to a series of conceptual logic troubles.

A question arises. All measured values that diverge from each other can definitely be used to describe a random variable, but a measured value must be a numerical value and belong to a constant. So, how should the measurement theory be interpreted?

So far, in the measurement industry, there is no literature questioning the conceptual category of measured values, such as the recent literature [6].

In the references [7,8,9,10], the authors proposed some new concepts to reinterpret measurement theory. Reference [7] proposes a new error epistemology that all errors follow a random distribution and cannot classified as systematic error and random error. Reference [8] points out, the standard deviation (variance) is the evaluation value of the probability interval of error, any error has a variance for evaluating its uncertainty, and so on. Reference [9] points out, the dispersion and deviation of repeated observations are determined by the changing rules of repeated measurement conditions, and it is possible and correct to handle errors according to the function model or the random model. Reference [10] points out, the measured value is a numerical value whose variance is zero, and the dispersion of a measured value is an incorrect concept. According to these new concepts, the mathematical expression of the Mount Everest elevation case should be $x = 8844.43\text{m}$ and $\sigma(\Delta x) = 0.21\text{m}$, where $x$ represents the measured value and $\Delta x$ represents its error.

Although these new concepts have been proposed to reinterpret measurement theory, the root of these concepts and the interpretation process have not been fully described mathematically. Therefore, in this paper, the authors follow strict mathematical concept to point out the misunderstanding of the traditional concepts, give a clear interpretation for the origin of these new measurement concepts, and systematically explain a series of changes in theoretical logic and mathematical processing.

## 2. Constant and random variable

In probability theory, a constant is a numerical value, such as 100, 150, $x$=100, $x$=8844.43, and so on.

Unlike constants, random variable is an unknown quantity whose actual value cannot be given. Because the random variable is unknown or uncertain, we can only describe the probability range of its value. In order to study its probability range, it is necessary to study the distribution range of all its possible values (sample space), while all possible values refer to the set of test values of random variables under all permitted possible test conditions (random test does not have the same conditions). Mathematical expectation and variance are the numerical expression of probability range of random variable.

For a random variable $L$ with all possible values $\{L_i\}$, there is $L \in \{L_i\}$, $P_i$ is the probability that each $L_i$ is $L$ (Continuous random variables correspond to the probability density function $P(L)$), and its mathematical expectation is defined as follows:

$$E(L) = \sum_{i=1}^{n} P_i L_i \quad \text{or} \quad E(L) = \int_{-\infty}^{+\infty} L P(L) dL \tag{2-1}$$

And its variance is defined as the dispersion of its possible values $\{L_i\}$:

$$\sigma^2(L) = E[L - E(L)]^2 \tag{2-2}$$

This means that the random variable $L$ exists within a probability interval with mathematical expectation $E(L)$ and variance $\sigma^2(L)$, or that mathematical expectation and variance are the evaluation values of its probability interval. Note that describing a random variable requires two parameters: mathematical expectation and variance, both of which are indispensable.

Now, suppose that there is a constant $C$, and there are $E(L) = C$ and $\sigma^2(L) = 0$, then:

$$E[L - E(L)]^2 = 0$$

By substituting $E(L) = C$, we get: $E(L-C)^2 = 0$

Therefore, $L = C$

That is, when the variance of a random variable $L$ is reduced to zero, it becomes a constant $C$. In other words, for a constant $C$, because all its possible values are itself, we get:

$$E(C) = C \tag{2-3}$$

$$\sigma^2(C) = E[C - E(C)]^2 = 0 \tag{2-4}$$

That is to say, a constant is a special random variable, and its mathematical expectation and variance are itself and 0 respectively. Of course, a constant is a known quantity, and usually does not need to be expressed in terms of probability.

It can be seen that both constant and random variable have their own mathematical expectation and variance. Therefore, if we can give the variance of a quantity but cannot give its mathematical expectation, there must be a conceptual mistake.

It should be noted that for the random variable $L \in \{L_i\}$, its basic feature is that its value is unknown; but for a sample $L_k \in \{L_i\}$, because it is a numerical value, it is still a constant rather than a random variable, and there are $E(L_k) = L_k$ and $\sigma^2(L_k) = 0$. Obviously, $E(L) \neq E(L_k)$ and $\sigma^2(L) \neq \sigma^2(L_k)$. That is, constant and random variable are distinguished by whether they have a numerical value, and the sample $L_k$is a numerical value, which is a constant and cannot be described by the entire set $\{L_i\}$. The conceptual differences between random variable and sample is shown in Table 1.

Table 1: The conceptual differences between the random variable and the sample.

| **Random variable $L$** | **Sample $L_k$** |
| --- | --- |
| $L \in \{L_i\}$ | $L_k \in \{L_i\}$ |
| The value of $L$ is unknown. | $L_k$ is a numerical value, which is a constant. |
| It has many possible values. | All its possible values are itself. |
| Every sample $L_i$ in $\{L_i\}$ is a possible value of $L$, that is, $P_i \neq 0\%$. | Any other samples $L_j$ in $\{L_i\}$ are definitely not $L_k$, that is, $P_j = 0\%$ and $P_k = 100\%$. |
| We can only use $\{L_i\}$ to describe its probability range. | It is a numerical value, and no probability range needs to be described. |
| $E(L) = \sum P_i L_i \quad \sigma^2(L) = E[L - E(L)]^2$ | $E(L_k) = L_k \quad \sigma^2(L_k) = 0$ |
| $E(L) \neq E(L_k) \qquad \sigma^2(L) \neq \sigma^2(L_k)$ | |

Example 1. A dice has six faces corresponding to the values 1, 2, 3, 4, 5 and 6 respectively. After the dice is thrown, we don't know its display value, which is a random variable$L \in \{1,2,3,4,5,6.\}$. Then, according to definitions (2-1) and (2-2), there are $E(L) = 3.5$ and $\sigma^2(L) = 2.92$. Its meaning is that although the display value $L$ is unknown, it exists within a probability interval with 3.5 as the center and 2.92 as the width evaluation. Obviously, there are

$$\left.\begin{array}{l} E(1) = 1 \\ E(2) = 2 \\ \vdots \\ E(6) = 6 \end{array}\right\} \neq E(L) = 3.5 \qquad \left.\begin{array}{l} \sigma^2(1) = 0 \\ \sigma^2(2) = 0 \\ \vdots \\ \sigma^2(6) = 0 \end{array}\right\} \neq \sigma^2(L) = 2.92$$

Example 2. The exam scores of all students in a school are $\{x_i\}$, in which the exam score of student A is $x_0$. It is reasonable to express an unknown score $x$ with the mathematical expectation $E(x)$ and variance $\sigma^2(x)$ of $\{x_i\}$, but it is illogical to impose $\sigma^2(x)$ on $x_0$, because the $x_0$ is a known constant and has no need to be expressed by variance and mathematical expectation. Moreover, $\sigma^2(x)$ is obviously not the dispersion of future exam scores of student A. That is $E(x) \neq E(x_0)$ and $\sigma^2(x) \neq \sigma^2(x_0)$.

Example 3. Someone's salary is $x_0 = 10000\text{RMB}$, and the salaries of all the employees in his company form a sample sequence $\{x_i\}$. By making the statistics of $\{x_i\}$, we can get the mathematical expectation $E(x)$ and variance$\sigma^2(x)$. In exactly the same way, although there is $x_0 \in \{x_i\}$, we cannot force $x_0$ to belong to random variable at all, because $x_0 = 10000\text{RMB}$ is known.

That is to say, it is reasonable to evaluate the probability of an unknown event with the statistic values of a group of known events, but it is illogical to use it to evaluate the "probability" of a known event.

Now, we suppose there is a random variable $L$ with mathematical expectation $E(L) = C$, and $\Delta C = L - E(L)$, then there is:

$$\begin{aligned} L &= E(L) + L - E(L) \\ &= C + \Delta C \end{aligned} \tag{2-5}$$

For the constant $C$, there are:

$$E(C) = C \tag{2-6}$$

$$\sigma^2(C) = 0 \tag{2-7}$$

For the random variable $\Delta C = L - E(L)$, there are:

$$\begin{aligned} E(\Delta C) &= E[L - E(L)] \\ &= E(L) - E(L) \\ &= 0 \end{aligned} \tag{2-8}$$

$$\begin{aligned} \sigma^2(\Delta C) &= E[\Delta C - E(\Delta C)]^2 \\ &= E(\Delta C^2) \\ &= E[L - E(L)]^2 \\ &= \sigma^2(L) \end{aligned} \tag{2-9}$$

This is to say, a random variable $L$ with mathematical expectation $C$ can be viewed as the superposition of a constant $C$ and a random variable $\Delta C$ taking 0 as its mathematical expectation. And it should be noted that the variance $\sigma^2(L)$ or $\sigma^2(\Delta C)$ always has nothing to do with constant $C$.

## 3. The origin of conceptual troubles in traditional theory

Figure 1 is the schematic diagram of the measurement concept in traditional measurement theory. Because people notice that measured value is in a state of random change in repeated measurement, the measured value and the random error are considered as random variables, and the variance is the dispersion of measured value or random error. Besides, the systematic error and the true value are constant in repeated measurement, so the systematic error and the true value are considered as constants which have no variances (or the variance is zero). In this way, according to formula (2-2), there is $\sigma^2(x) = E[x - E(x)]^2$. Therefore, traditional textbooks[11,12,13] usually use the form of $\sigma^2(x)$ or $\sigma_x^2$ to express the variance. However, these are obviously inconsistent with the meanings of random variables and constants described in Section 2.

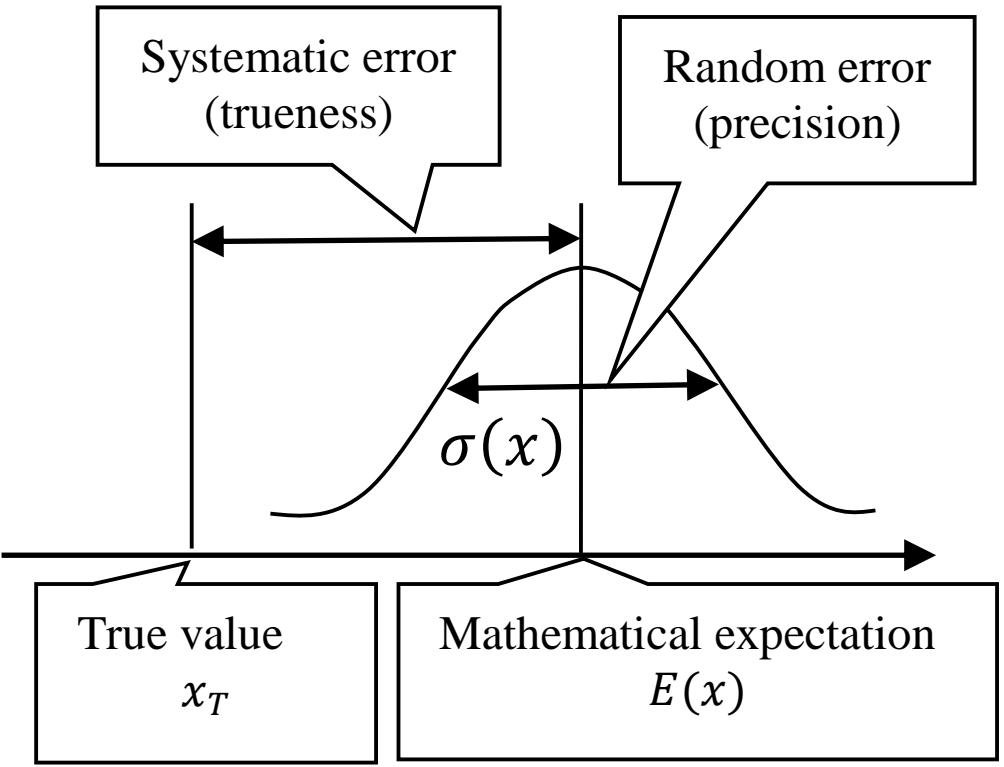

**Figure 1. Schematic diagram in traditional measurement theory**

Besides, in actual measurement, we always have to give a numerical value $x_0$ as the final measured value. Therefore, the actual schematic diagram is shown in Figure 2. According to the concepts in section 2, although measured value $x_0$ is a sample within a random distribution, because the measured value $x_0$ is a numerical value and there are $\sigma^2(x_0) = 0$ and $E(x_0) = x_0$, it is illogical to replace $\sigma^2(x_0)$ with $\sigma^2(x)$.

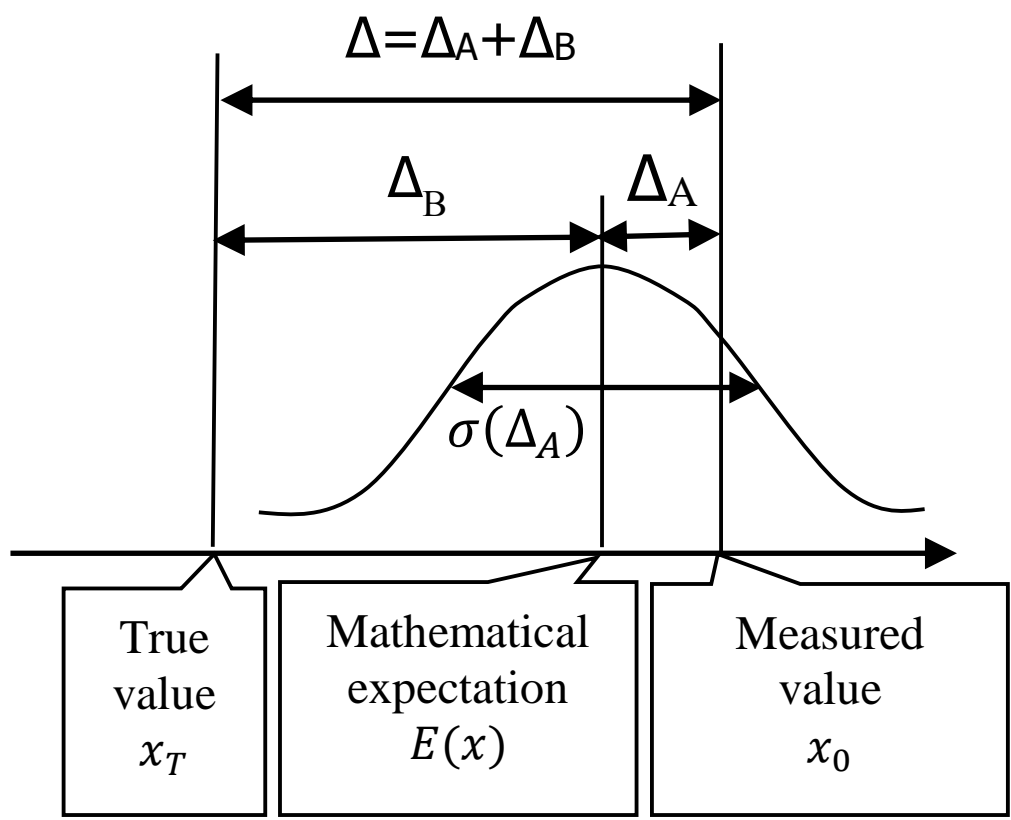

**Figure 2. Schematic diagram in the new concept theory**

In addition, simply replacing $\sigma^2(x_0)$ with $\sigma^2(x)$ cannot express a complete mathematical meaning, because the traditional theory cannot submit the mathematical expectation $E(x)$.

On the other hand, the systematic error and the true value are unknown and are regarded as constants by traditional theory. However, according to formula（2-3）, the mathematical expectation of a constant is itself, so it is impossible to give the numerical values of mathematical expectations of systematic error and true value. Therefore, this so-called constant is obviously not the same concept as the constant in the probability theory, and it is also a conceptual trouble in the traditional measurement theory.

In short, in the traditional theory, except for the conceptual trouble of violating the concept that the variance of a constant is zero, the conceptual trouble of missing mathematical expectations is shown in Table 2.

Table 2: The conceptual trouble of missing mathematical expectations.

| | **Measured value** $\boldsymbol{x}$ | **Random error** $\boldsymbol{x - E(x)}$ | **Systematic error** $\boldsymbol{E(x) - x_T}$ | **True value** $\boldsymbol{x_T}$ |
|---|---|---|---|---|
| Mathematical expectation | Absent | 0 | Absent | Absent |
| Variance | $\sigma^2(x)$ | $\sigma^2(x)$ | 0 | 0 |

Because random errors have variance but systematic errors cannot be quantitatively evaluated, traditional theories believe that trueness and accuracy are both qualitative concepts, and the relationship between the uncertainty concept and them is of course very difficult to explain.

## 4. Probability expression of basic measurement concepts

It can be seen that because of the wrong understanding of the concept of random variable, the conceptual logic of traditional theory actually has systematic troubles. Therefore, we need to reorganize the basic measurement concept logic according to the concepts in section 2.

### 4.1 Measured value

In Figure 2, the measured value $x_0$ is an observed value within all possible observed values $\{x_i\}$, and is a numerical value. According to formulas (2-3) and (2-4), there are:

$$E(x_0) = x_0 \tag{4-1}$$

$$\sigma^2(x_0) = 0 \tag{4-2}$$

Please note that any quantity with a numerical value, including the error sample, the measured value of error, the detected value of instrument error, the value of mathematical expectation, the value of variance, and so on, are constants.

### 4.2 Error

As shown in Figure 2, the true value of measurand is $x_T$, and the error of the final measured value $x_0$ , which is a unknown deviation $\Delta = x_0 - x_T$ , can be divided into $\Delta_A = x_0 - E(x)$ and $\Delta_B = E(x) - x_T$.

First of all, the error $\Delta_A$ is a random variable, and there is $\Delta_A \in \{x_i - E(x)\}$. Moreover, the sample space of error $x - E(x)$ is also $\{x_i - E(x)\}$, so error $x - E(x)$ can be used to represent $\Delta_A$, that is, $\Delta_A = x - E(x)$. According to the formulas (2-1) and (2-2), there are:

$$\begin{aligned} E(\Delta_A) &= E[x - E(x)] \\ &= 0 \end{aligned} \tag{4-3}$$

$$\begin{aligned} \sigma^2(\Delta_A) &= E\{x - E(x) - E[x - E(x)]\}^2 \\ &= E[x - E(x)]^2 \\ &= E(\Delta_A{}^2) \end{aligned} \tag{4-4}$$

That is, although deviation $\Delta_A$ is unknown, it exists within a probability interval with 0 as center and $\sigma^2(\Delta_A)$ as width evaluation. In other words, $\sigma^2(\Delta_A)$ is the evaluation of probability interval of deviation $\Delta_A$, which expresses the degree that surveyor cannot determine the value of deviation $\Delta_A$.

Taking the normal distribution as an example, the variance $\sigma^2(\Delta_A)$ expresses that the deviation $\Delta_A$ is within the interval of $[-\sigma(\Delta_A), +\sigma(\Delta_A)]$ under the confidence probability of 68%. Variance is actually a concept of error range with probability meaning, and expresses an error's possible deviation degree.

In formula (4-4), the single deviation $\Delta_A$ is a member within all its possible values, and the dispersion interval of all its possible values is the probability interval of this deviation $\Delta_A$. An unknown deviation follows a random distribution, which means that all possible values of the deviation follow a random distribution.

This principle obviously can be extended to the error $\Delta_B$. In fact, when we trace back to the upstream measurement of forming error $\Delta_B$, we will find that the formation principle of error $\Delta_B$ is similar to that of current error $\Delta_A$, and that the error $\Delta_B$ is also a member within all its possible values. Therefore, there is also a variance $\sigma^2(\Delta_B)$ to evaluate the probability interval of error $\Delta_B$, and there is also $E(\Delta_B) = 0$.

For example, the multiplicative constant error R of a geodimeter [14,15] comes from the frequency error of the quartz crystal, and is always viewed as a systematic error without variance by traditional measurement theory. However, it is the output error in the field of instrument manufacturing, and the submission process of its variance will be demonstrated in the case in Section 6.

Obviously, according to the principle of formula (2-5) ~ (2-9), if the mathematical expectation of an error is C rather than 0, then C must be corrected to the final measured value, and the mathematical expectation of the remaining unknown error is still 0. That is, for any unknown error $\Delta x$, there is always

$$E(\Delta x) = 0 \tag{4-5}$$

Thus, the error's variance is expressed as below:

$$\begin{aligned} \sigma^2(\Delta x) &= E[x - E(\Delta x)]^2 \\ &= E(\Delta x^2) \end{aligned} \tag{4-6}$$

The $\Delta x$ in formulas (4-5) and (4-6) can express not only the deviation $\Delta_A$ between the measured value and the mathematical expectation, but also the deviation $\Delta_B$ between the mathematical expectation and the true value. It can even express the deviation $\Delta = \Delta_A + \Delta_B$ between the measured value and the true value.

In this way, according to formulas (2-1) and (2-2), there are:

$$\Delta = \Delta_A + \Delta_B \tag{4-7}$$

$$\begin{aligned} E(\Delta) &= E(\Delta_A) + E(\Delta_B) \\ &= 0 \end{aligned} \tag{4-8}$$

$$\begin{aligned} \sigma^2(\Delta) &= E[\Delta - E(\Delta)]^2 \\ &= E(\Delta^2) \\ &= E(\Delta_A + \Delta_B)^2 \\ &= \sigma^2(\Delta_A) + \sigma^2(\Delta_B) \end{aligned} \tag{4-9}$$

Because the final measured value is unique and constant, both $\Delta_A$ and $\Delta_B$ are unknown deviations. In addition, both of them have their own variance, hence, it is incorrect that the traditional measurement theory considers $\Delta_A$ as random error and considers $\Delta_B$ as systematic error. Moreover, the corresponding concepts of precision and trueness are also incorrect.

It should be emphasized that, the formula (4-5) means that the mean value of all possible values of an unknown error is 0, which expresses the probabilities that an unknown error takes positive and negative value are equal in our subjective cognition. From a statistical perspective, all possible values of an error refer to the set of all error values under all possible measurement conditions permitted by measurement specification, so the traditional concept of "repeated measurement under the same conditions" must be abandoned [8,9], otherwise an unique error value obtained under a particular condition is only one sample within all possible values and does not represent all possible values, which is very easy to cause the illusion of $E(\Delta x) \neq 0$.

In short, being different from measured value, the error is unknown; because the error is unknown, we can only study its probability range; because of studying its probability range, we must study all possible values of error; because of studying all possible values of error, error samples must come from all the possible measurement conditions permitted by measurement specification, and the traditional concept of "repeated measurement under the same measurement conditions" must be abandoned.

### 4.3 Variance of regular error

Any error has its variance, including the regular error, because the regular error also has all its possible values.

For example, the periodic error of a phase photoelectric distance meter [14,15] conforms to the sine regularity, and its function model is $\delta = A\sin(\frac{D}{\lambda} \times 2\pi + \phi)$. However, when we only observe the density distribution of all its possible values, this cyclic error's probability density function $f(\delta)$ can be derived as:

$$f(\delta) = \begin{cases} \dfrac{1}{\pi\sqrt{A^2 - \delta^2}} & \left(|\delta| \le A\right) \\ 0 & \left(|\delta| > A\right) \end{cases}$$

**Figure 3. Regularity and randomness of periodic error**

As shown in Figure 3, further, its variance can be derived as $\sigma^2(\delta) = \frac{A^2}{2}$, and its mathematical expectation can be derived as $E(\delta) = 0$.

Another example, the rounding error $\delta$ is a sawtooth cycle regularity function of true value $w$. However, when we only observe the density distribution of all possible values, the error also follows a random distribution, as shown in Figure 4, and its probability density function is:

$$f(\delta) = \begin{cases} \dfrac{1}{2a} & \left(|\delta| \le a\right) \\ 0 & \left(|\delta| > a\right) \end{cases}$$

**Figure 4. Regularity and randomness of rounding error**

Its variance can be derived as $\sigma^2(\delta)=\frac{a^2}{3}$, and its mathematical expectation can be derived as $E(\delta)=0$.

That is to say, the regularity and the randomness are the effect of observing all possible values of error from different perspectives, there is no opposition between them, and there is actually no need to dwell on the error's regularity in the discussion of error evaluation. In other words, when a regular error is unknown, we can still use the mathematical expectation and the variance to describe its probability range. Furthermore, traditional measurement theories use the regularity and the randomness to classify errors into systematic errors and random errors, which is also proved to be incorrect.

In addition, when the measurement conditions vary with the repeated measurement, the corresponding regular errors will cause the dispersion of the repeated observations, which is exactly the same as the dispersion caused by the noise which vary with time conditions[9]. This dispersion is exactly the means by which we obtain its variance. It can be seen that, in order to obtain the variance of an error, we need to collect error samples under all possible measurement conditions.

### 4.4 Probability expression of true value

With previous statements we already know that for the measured value $x_0$, there are $E(x_0)=x_0$ and $\sigma^2(x_0)=0$; for the error $\Delta$, there are $E(\Delta)=0$ and $\sigma^2(\Delta)=E(\Delta)^2$. Because the error is the difference between the measured value and the true value, that is $\Delta=x_0-x_T$, there are:

$$x_T=x_0-\Delta \tag{4-10}$$

$$\begin{aligned} E(x_T) &= E(x_0-\Delta) \\ &= E(x_0)-E(\Delta) \\ &= x_0 \end{aligned} \tag{4-11}$$

$$\begin{aligned} \sigma^2(x_T) &= E(x_T-Ex_T)^2 \\ &= E(x_0-\Delta-x_0)^2 \\ &= \sigma^2(\Delta) \end{aligned} \tag{4-12}$$

The probability expressions of the true value $x_T$, the measured value $x_0$ and the error $\Delta$ are summarized in Table 3[10].

Table 3: The probability expression of true value, measured value and error.

| | **Measured value $x_0$** | **Error $\Delta$** | **True value $x_T$** |
|---|---|---|---|
| Mathematical expectation | $x_0$ | 0 | $x_0$ |
| Variance | 0 | $\sigma^2(\Delta)$ | $\sigma^2(\Delta)$ |

The above is the case where an observed value is used as the final measured value. If the mean value of $n$ observations is taken as the final measured value, it can be inferred that the variance $\sigma^2(\Delta_A)$ will decrease by $n$ times. Please see section 6.1.

## 5. Covariance propagation

It can be seen from Table 3 that after a measured value $x_0$ is given, only the variance $\sigma^2(\Delta)$ needs to be studied.

### 5.1 Covariance

Considering that two correlated errors have common component, the formula (4-6) can be extended to any two errors. That is

$$\sigma(\Delta x_i \Delta x_j) = E(\Delta x_i \Delta x_j) \tag{5-1}$$

Thus, for the error sequence $\Delta\mathbf{X} = (\Delta x_1 \quad \Delta x_2 \quad \cdots \quad \Delta x_t)^T$, the definition of variance is:

$$\mathbf{D}(\Delta\mathbf{X}) = E(\Delta\mathbf{X})(\Delta\mathbf{X})^T \tag{5-2}$$

That is

$$\mathbf{D}(\Delta\mathbf{X}) = E\begin{pmatrix} \Delta x_1 \\ \Delta x_2 \\ \vdots \\ \Delta x_t \end{pmatrix}\begin{pmatrix} \Delta x_1 & \Delta x_2 & \cdots & \Delta x_t \end{pmatrix}$$

$$= \begin{pmatrix} \sigma_{11}^2 & \sigma_{12} & \cdots & \sigma_{1t} \\ \sigma_{21} & \sigma_{22}^2 & \cdots & \sigma_{2t} \\ \vdots & \vdots & & \vdots \\ \sigma_{t1} & \sigma_{t2} & \cdots & \sigma_{tt}^2 \end{pmatrix}$$

Obviously, the true definition of the variance is formula (5-2). Formula (4-6) is only a special case of formula (5-2) when $t$=1, and formula (4-4) is only a special case of formula (4-6) when interpreting $\Delta x$ as $x - E(x)$.

So, what is the meaning of covariance?

It is assumed that the errors $k$, $p$ and $q$ are uncorrelated with each other, and that their variance are $\sigma_k^2$, $\sigma_p^2$ and $\sigma_q^2$ respectively. Now, there are two errors $\delta = k + p$ and $\varepsilon = k + q$, and they contain a communal error component $k$. Therefore, we can get:

$$\sigma_\delta^2 = \sigma_k^2 + \sigma_p^2 \tag{5-3}$$

$$\sigma_\varepsilon^2 = \sigma_k^2 + \sigma_q^2 \tag{5-4}$$

According to the definition of covariance:

$$\begin{aligned} \sigma_{\delta\varepsilon} &= E(\delta\varepsilon) \\ &= E[(k+p)(k+q)] \\ &= E(k^2) + E(kp) + E(kq) + E(pq) \end{aligned} \tag{5-5}$$

With the assumption that the errors $k$, $p$ and $q$ are irrelevant from each other, we can get $E(kp) = 0$, $E(kq) = 0$ and $E(pq) = 0$, and equation (5-5) becomes to:

$$\sigma_{\delta\varepsilon} = E(k^2) = \sigma_k^2 \tag{5-6}$$

The covariance $\sigma_{\delta\varepsilon}$ is actually the variance of their communal error component $k$. That is to say, the mathematical meaning of covariance is the probability evaluation of the communal error component contained in two errors. As long as there are communal error component among different errors, there must be a covariance between them. Of course, the symbol and coefficient of communal error component should be considered in the actual measurement.

For example, the two measured value's errors measured by the same instrument are correlated, and the errors of two instruments calibrated by the same benchmark are also correlated.

Moreover, like the above principle, when two errors are associated with the same measurement condition, there is also a covariance between them. For example, both the error of light speed in atmosphere and the thermal expansion error of metal are functions of temperature, and there is a correlation between all possible values of the two errors.

### 5.2 The law of covariance propagation

Because any error has all its possible value and has its variance, the law of covariance propagation is extended to any error. In addition, the law of covariance propagation can only be interpreted as the propagation law of error's probability interval, and cannot be interpreted as propagation law of measured value's dispersion.

Here is a measurement equation: $\mathbf{Z} = F(\mathbf{X})$ (5-7)

We can derive the total differential of Equation (5-7):

$$\Delta\mathbf{Z} = \mathbf{K} \cdot \Delta\mathbf{X} \tag{5-8}$$

Where $\Delta\mathbf{Z} = \begin{pmatrix} \Delta Z_1 \\ \Delta Z_2 \\ \vdots \\ \Delta Z_t \end{pmatrix}$, $\mathbf{K} = \begin{pmatrix} k_{11} & k_{12} & \cdots & k_{1n} \\ k_{21} & k_{22} & \cdots & k_{2n} \\ \vdots & \vdots & & \vdots \\ k_{t1} & k_{t2} & \cdots & k_{tn} \end{pmatrix}$, $\Delta\mathbf{X} = \begin{pmatrix} \Delta x_1 \\ \Delta x_2 \\ \vdots \\ \Delta x_n \end{pmatrix}$.

According to formula (5-2), the covariance matrix of the error sequence $\Delta\mathbf{Z}$ is

$$\begin{aligned} \mathbf{D}(\Delta\mathbf{Z}) &= E(\Delta\mathbf{Z})(\Delta\mathbf{Z})^{\mathrm{T}} \\ &= E(\mathbf{K} \cdot \Delta\mathbf{X})(\mathbf{K} \cdot \Delta\mathbf{X})^{\mathrm{T}} \\ &= \mathbf{K} \cdot \mathbf{D}(\Delta\mathbf{X}) \cdot \mathbf{K}^{\mathrm{T}} \end{aligned} \tag{5-9}$$

Equation (5-9) is the law of covariance propagation. The relationship between equations (5-8) and (5-9) is:

1. In the error equation (5-8), the direct participants of synthesis is the error itself, each error is a deviation, and error synthesis always follows the algebra rule.

2. In the variance equation (5-9), the participants of synthesis are all possible values of each error instead of each error itself. It expresses the propagation relation of dispersion of all possible values of errors, and also the propagation relation of probability intervals between errors.

## 6. Statistical calculation of variance

Since the number of error samples is always limited in actual measurement, formula (4-6) can be approximated as

$$\sigma^2(\Delta x) \approx \frac{\sum_{i=1}^{n} (\Delta x_i)^2}{n} \tag{6-1}$$

Formula (6-1) is also the source of least squares principle. That is to say, from the perspective of the new concept, only the concept of error evaluation changes, while the principle of the least square method used to obtain the best measured values does not change.

In actual measurement, in order to achieve the reduction and evaluation of the measurement error, a large number of observations should be carried out. Because errors make a large number of observations contradict from each other, the optimal measured values must be given by adjustment process, and the errors of the measured values should be also evaluated. We only discuss the case of the least square adjustment in this section.

### 6.1 Direct measurement for single measurand

A measurand is directly measured by $n$ times, and $n$ observations $x_i$ are obtained. In this way, using $y_0$ to represent the best measured value, the error equations are:

$$\left.\begin{aligned} v_1 &= x_1 - y_0 \\ v_2 &= x_2 - y_0 \\ &\vdots \\ v_n &= x_n - y_0 \end{aligned}\right\} \tag{6-2}$$

According to the least square method, the final measured value is

$$y_0 = \sum_{i=1}^{n} \frac{x_i}{n} \tag{6-3}$$

The measurement model of this measurement method is $V = X - Y$, $v_i$, $x_i$ and $y_0$ are samples of random variables $V$, $X$ and $Y$ respectively.

Taking the total differential of equation (6-3), the error propagation equation is

$$\Delta y = \sum_{i=1}^{n} \frac{\Delta x_i}{n} \tag{6-4}$$

Now we only discuss the variances of the error components $x_i - E(X)$ and $y_0 - E(Y)$, and make $\Delta x_i = x_i - E(X)$ and $\Delta y = y_0 - E(Y)$.

Because $\Delta x_i = x_i - E(X)$ is unknown, is a random variable, and has the same sample space $\{x_i - E(X)\}$ as $X - E(X)$, so $\Delta x_i = x_i - E(X)$ can be represented by $\Delta x = X - E(X)$, that is, $\Delta x_i = \Delta x = X - E(X)$. Similarly, there is $\Delta y = Y - E(Y)$.

By applying the law of covariance propagation to formula (6-4), there is,

$$\sigma^2(\Delta y) = \frac{\sigma^2(\Delta x)}{n} \tag{6-5}$$

According to the measurement model $V = X - Y$, there is

$$\begin{aligned} \Delta x &= X - E(X) \\ &= V + Y - E(V + Y) \\ &= V + \Delta y \end{aligned} \tag{6-6}$$

Therefore, according to the definition of variance, there is

$$\begin{aligned} \sigma^2(\Delta x) &= E[(\Delta x)^2] \\ &= E[V + \Delta y]^2 \\ &= E(V)^2 + E(\Delta y)^2 \\ &= E(V)^2 + \sigma^2(\Delta y) \end{aligned} \tag{6-7}$$

Substituting $E(V)^2 \approx \frac{1}{n}\sum_{i=1}^{n} v_i^2$ and formula (6-5) into equation (6-7), we get:

$$\sigma^2(\Delta x) \approx \frac{\sum_{i=1}^{n} v_i^2}{n} + \frac{\sigma^2(\Delta x)}{n} \tag{6-8}$$

Therefore

$$\sigma(\Delta x) \approx \sqrt{\frac{\sum_{i=1}^{n} v_i^2}{n-1}} \tag{6-9}$$

### 6.2 Indirect measurement for single measurand

Different from the direct measurement, each observation $x_i$ in the indirect measurement is the measured data of $a_i$ times of measurand. The error equations of the repeated measurement are:

$$\left.\begin{aligned} v_1 &= x_1 - a_1 y_0 \\ v_2 &= x_2 - a_2 y_0 \\ &\vdots \\ v_n &= x_n - a_n y_0 \end{aligned}\right\} \tag{6-10}$$

According to the least square method, the final measured value is:

$$y_0 = \frac{\sum_{i=1}^{n} a_i x_i}{\sum_{i=1}^{n} a_i^2} \tag{6-11}$$

The measurement model is $V_i = X_i - a_i Y$, $v_i$, $x_i$ and $y_0$ are samples of random variables $V_i$, $X_i$ and $Y$ respectively.

Similarly, for the errors $\Delta x = X_i - E(X_i)$ and $\Delta y = Y - E(Y)$, the covariance propagation relationship is

$$\sigma^2(\Delta y) = \frac{\sigma^2(\Delta x)}{\sum_{i=1}^{n} a_i^2} \tag{6-12}$$

Similarly, according to the measurement model $V_i = X_i - a_i Y$, there is

$$\begin{aligned} X_i - E(X_i) &= V_i + a_i Y - E(V_i + a_i Y) \\ &= V_i + a_i \Delta y \end{aligned} \tag{6-13}$$

According to the definition of variance, there is

$$\begin{aligned} \sigma^2(\Delta x) &= E[\Delta x]^2 \\ &= \lim_{n\to\infty} \frac{1}{n} \{[X_1 - E(X_1)]^2 + [X_2 - E(X_2)]^2 + \cdots\} \\ &= \lim_{n\to\infty} \frac{1}{n} [(V_1 + a_1 \Delta y)^2 + (V_2 + a_2 \Delta y)^2 + \cdots] \\ &\approx \frac{1}{n} \sum_{i=1}^{n} v_i^2 + \frac{\sum_{i=1}^{n} a_i^2}{n} \sigma^2(\Delta y) \\ &= \frac{1}{n} \sum_{i=1}^{n} v_i^2 + \frac{\sigma^2(\Delta x)}{n} \end{aligned} \tag{6-14}$$

Therefore

$$\sigma(\Delta x) \approx \sqrt{\frac{\sum_{i=1}^{n} v_i^2}{n - 1}} \tag{6-15}$$

### 6.3 Indirect measurement for multiple measurands

In this measurement mode, there are $t$ different measurands, and each observation value $x_i$ is obtained by measuring the linear superposition value of multiple measurands. The error equations of the repeated measurement are:

$$\begin{pmatrix} v_1 \\ v_2 \\ \vdots \\ v_n \end{pmatrix} = \begin{pmatrix} x_1 \\ x_2 \\ \vdots \\ x_n \end{pmatrix} - \begin{pmatrix} a_{11} & a_{12} & \cdots & a_{1t} \\ a_{21} & a_{22} & \cdots & a_{2t} \\ & & \vdots & \\ a_{n1} & a_{n2} & \cdots & a_{nt} \end{pmatrix} \begin{pmatrix} y_1 \\ y_2 \\ \vdots \\ y_t \end{pmatrix} \tag{6-16}$$

That is:

$$\mathbf{V} = \mathbf{X} - \mathbf{AY} \tag{6-17}$$

According to the principle of the least squares, its measured values are:

$$\mathbf{Y} = \left[\mathbf{A}^T\mathbf{A}\right]^{-1}\mathbf{A}^{\mathrm{T}}\mathbf{X} \tag{6-18}$$

The measurement model is $V_i = X_i - \begin{pmatrix} a_{i1} & a_{i2} & \cdots & a_{it} \end{pmatrix} \begin{pmatrix} Y_1 \\ Y_2 \\ \vdots \\ Y_t \end{pmatrix}$, $v_i$, $x_i$ and $y_j$ are samples of random variables $V_i$, $X_i$ and $Y_j$ respectively.

The error propagation equation is:

$$\Delta\mathbf{Y} = \left[\mathbf{A}^T\mathbf{A}\right]^{-1}\mathbf{A}^{\mathrm{T}}\Delta\mathbf{X} \tag{6-19}$$

Similarly, for the errors $\Delta x = X_i - E(X_i)$ and $\Delta y_j = Y_j - E(Y_j)$, the covariance propagation relationship is:

$$\mathbf{D}(\Delta\mathbf{Y}) = \sigma^2(\Delta x)\left[\mathbf{A}^T\mathbf{A}\right]^{-1} \tag{6-20}$$

Similarly, according to the measurement model, there is

$$X_i - E(X_i) = V_i + \begin{pmatrix} a_{i1} & a_{i2} & \cdots & a_{it} \end{pmatrix} \begin{pmatrix} \Delta y_1 \\ \Delta y_2 \\ \vdots \\ \Delta y_t \end{pmatrix} \tag{6-21}$$

According to the definition of variance, there is

$$\begin{aligned} \sigma^2(\Delta x) &= E(\Delta x)^2 \\ &= \lim_{n\to\infty} \frac{1}{n}\{[X_1 - E(X_1)]^2 + [X_2 - E(X_2)]^2 + \cdots\} \\ &= \lim_{n\to\infty} \frac{1}{n}\sum_{i=1}^{n}\left[V_i + (a_{i1} + a_{i2} + \ldots + a_{it})\begin{pmatrix} \Delta y_1 \\ \Delta y_2 \\ \vdots \\ \Delta y_t \end{pmatrix}\right]^2 \\ &\approx \frac{1}{n}\sum_{i=1}^{n} v_i^2 + \frac{1}{n}\sum_{i=1}^{n}\left[\begin{pmatrix} \Delta y_1 & \Delta y_2 & \cdots & \Delta y_t \end{pmatrix}\mathbf{A}^T\mathbf{A}\begin{pmatrix} \Delta y_1 \\ \Delta y_2 \\ \vdots \\ \Delta y_t \end{pmatrix}\right] \end{aligned} \tag{6-22}$$

Omitting the tedious algebraic calculation process, the final result is

$$\sigma(\Delta x) \approx \sqrt{\frac{\sum_{i=1}^{n} v_i^2}{n-t}} \tag{6-23}$$

As you can see, the change in Bessel's formula is that $\sigma(x)$ is written as $\sigma(\Delta x)$, and $\sigma(\Delta x)$ represents the dispersion of all possible values of the deviation $\Delta x_i = x_i - E(X)$. Also, the standard deviation $\sigma(\Delta y)$ or $\sigma(\Delta y_j)$, which is given by the formula (6-5), (6-12) or (6-20), is also the evaluation of probability interval of the deviation $\Delta y = y_0 - E(Y)$ or $\Delta y_j = y_j -$ $E(Y_j)$. Obviously, it is incorrect to express Bessel's formula as $\sigma^2(x_k) \approx \frac{1}{n-1}\sum_{j=1}^{n}\left(x_j - \bar{x}\right)^2$ in the literatures[1,4], because $x_k$ is a numerical value.

For example: the measured frequency values of a quartz crystal at different temperatures are shown in Table 4, and the nominal value of frequency is $f_0$ =5.000050MHz. Now, we need to give a temperature correction model for the frequency error and evaluate the standard deviation of the residual error after correction.

Table 4: Observed values

| **Temperature ˚C** | **Frequency MHz** | **Error value** $R_i = \Delta f_i / f_0 (1\times 10^{-6})$ |
|---|---|---|
| -40 ° | 4.999900 | -30 |
| -30 ° | 4.999975 | -15 |
| -20 ° | 5.000040 | -2 |
| -10 ° | 5.000085 | 7 |
| 0 ° | 5.000115 | 13 |
| 10 ° | 5.000110 | 12 |
| 20 ° | 5.000070 | 4 |
| 30 ° | 5.000035 | -3 |
| 40 ° | 5.000010 | -8 |
| 50 ° | 4.999995 | -11 |
| 60 ° | 4.999995 | -11 |
| 70 ° | 5.000010 | -8 |
| 80 ° | 5.000045 | -1 |
| 90 ° | 5.000125 | 15 |
| 100 ° | 5.000235 | 37 |

We use the first 4 terms of the Taylor series as the temperature model of the frequency error, that is $R = a + bT + cT^2 + dT^3$.

In this way, the error equation set is:

$$\begin{pmatrix} v_1 \\ v_2 \\ \vdots \\ v_n \end{pmatrix} = \begin{pmatrix} R_1 \\ R_2 \\ \vdots \\ R_n \end{pmatrix} - \begin{pmatrix} 1 & T_1 & T_1^2 & T_1^3 \\ 1 & T_2 & T_2^2 & T_2^3 \\ \vdots & \vdots & \vdots & \vdots \\ 1 & T_n & T_n^2 & T_n^3 \end{pmatrix} \begin{pmatrix} a \\ b \\ c \\ d \end{pmatrix}$$

According to the least square method, there is

$$\begin{pmatrix} n & \sum T_i & \sum T_i^2 & \sum T_i^3 \\ \sum T_i & \sum T_i^2 & \sum T_i^3 & \sum T_i^4 \\ \sum T_i^2 & \sum T_i^3 & \sum T_i^4 & \sum T_i^5 \\ \sum T_i^3 & \sum T_i^4 & \sum T_i^5 & \sum T_i^6 \end{pmatrix} \begin{pmatrix} a \\ b \\ c \\ d \end{pmatrix} = \begin{pmatrix} \sum R_i \\ \sum R_i T_i \\ \sum R_i T_i^2 \\ \sum R_i T_i^3 \end{pmatrix}$$

Substituting the values in Table 4 into above equation, there are:

$$\begin{pmatrix} 15 & 450 & 41500 & 292500 \\ 450 & 41500 & 2925000 & 256870000 \\ 41500 & 2925000 & 256870000 & 21952500000 \\ 292500 & 256870000 & 21952500000 & 1983295000000 \end{pmatrix} \begin{pmatrix} a \\ b \\ c \\ d \end{pmatrix} = \begin{pmatrix} -1 \\ 4610 \\ 304500 \\ 42713000 \end{pmatrix}$$

Solving the equations, get:

$$a = 9.983251, b = -0.013518, c = -0.018601, d = 0.000214.$$

Therefore, the frequency error's function model is fitted as:

$$R = 9.983251 - 0.013518T - 0.018601T^2 + 0.000214T^3$$

Fig5 is the comparison curve between the model and the actual error. According the formula (6-23), the standard deviation of residual error is

$$\sigma(\Delta R) = \sqrt{\frac{\sum_{i=1}^{n} v_i^2}{n-4}} = \pm 2.3 \times 10^{-6}$$

Finally, the frequency of quartz crystal is given as follows:

$$f = f_0(1 + R \times 10^{-6})$$

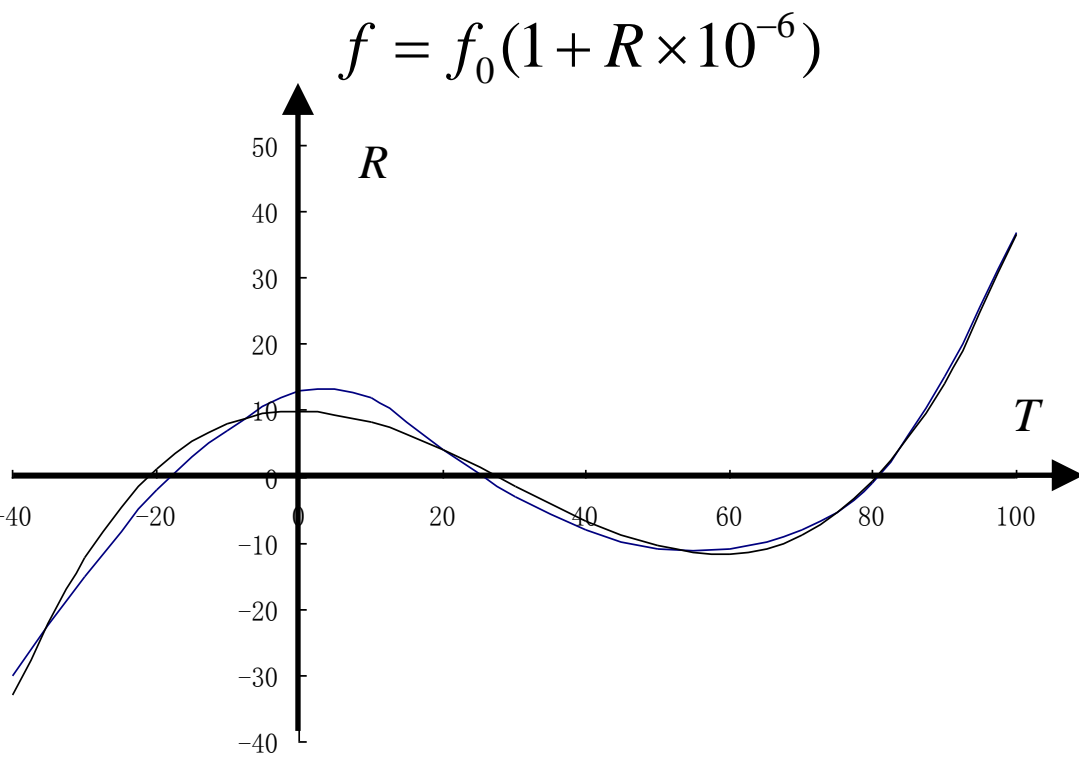

**Figure5.The function model fitting of frequency error**

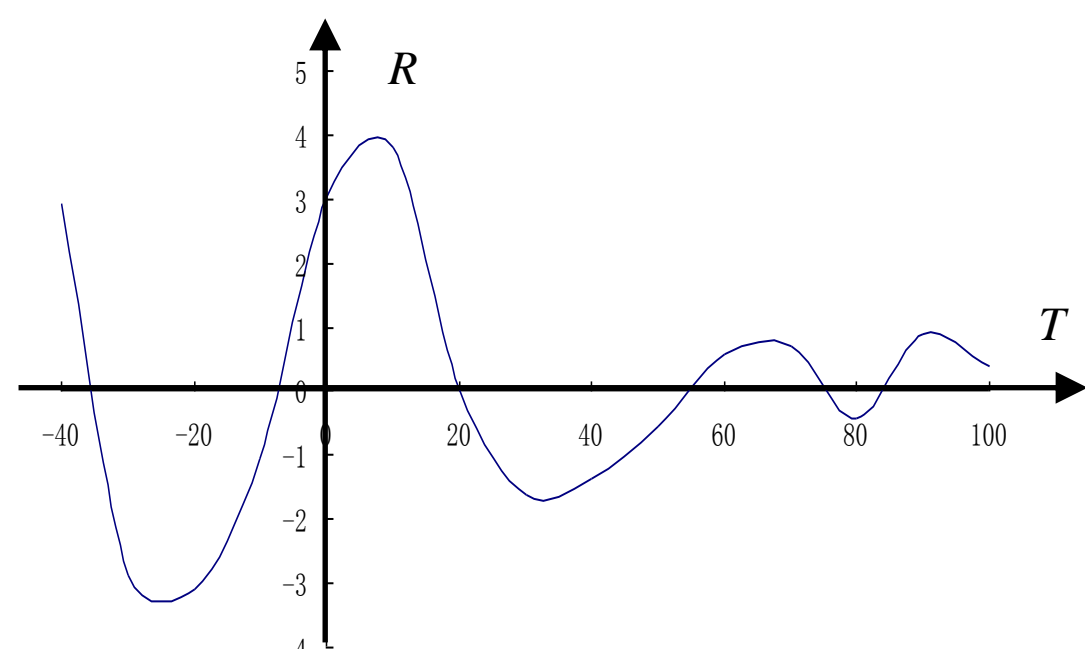

**Figure6.The residual error's curve**

That is, temperature-frequency error can be corrected by the measured values of temperature sensor, and a more accurate frequency value can be calculated. Residual error (as shown in Fig6), which is still a regular error, is also processed by statistical rules, and the standard deviation of the residual error is $\pm 2.3 \times 10^{-6}$. This error processing method has been widely used in the manufacture of photoelectric geodimeter [14,15].

## 7. Uncertainty

According to Figure 2, the total error of the final measured value is

$$\Delta = \Delta_A + \Delta_B \tag{7-1}$$

Where $\Delta_A$ is the deviation between measured value and expectation, and $\Delta_B$ is the deviation between expectation and true value.

Because the two errors are usually irrelevant, according the law of covariance propagation (5-9), there is:

$$\sigma(\Delta) = \sqrt{\sigma^2(\Delta_A) + \sigma^2(\Delta_B)} \tag{7-2}$$

This total standard deviation $\sigma(\Delta)$ is the evaluation of probability interval of total error $\Delta$ (the dispersion of all possible values of total error $\Delta$). It can be seen that formula (7-2) is consistent with the traditional uncertainty evaluation (But the expression is changed from $\sigma(x)$ to $\sigma(\Delta)$). Therefore, this total standard deviation $\sigma(\Delta)$ is actually the uncertainty, which expresses the probability range of the total error of final measured value. And the uncertainty concept, which is interpreted as the dispersion of measured value (constant) in the traditional measurement theory, is also proved to be incorrect.

It can be seen from Table 3 that the uncertainty is also the possible degree that the true value deviates from the measured value. That is, the uncertainty is not only the uncertainty of the error but also the uncertainty of the true value, but is not the uncertainty of the measured value. The measured value, which is a numerical value, has no uncertainty.

According to the interpretation of the traditional theory, $\sigma(\Delta_A)$ and $\sigma(\Delta_B)$ are referred as the uncertainty of Type A and the uncertainty of Type B respectively. However, the current $\sigma(\Delta_B)$ is actually the $\sigma(\Delta)$ of historical upstream measurement, and the current $\sigma(\Delta)$ can also be used as the $\sigma(\Delta_B)$ in future downstream measurement. This kind of interpretation with A/B classification of the uncertainty evaluation is obviously too rigid.

Moreover, the currently widely used formula (7-2) is only applicable to the direct repeated measurement model in section 6.1, but has no use at all for the indirect repeated measurement in sections 6.2 and 6.3, because in indirect repeated measurement, there are usually some error sources which not only contribute to dispersion of repeated observations but also contribute to their deviation, and it is difficult to distinguish them with A/B classification method. Therefore, A/B classification method is not universal in practice.

Formula (7-2) comes from the covariance propagation law (5-9). Thus, the basic principle of uncertainty synthesis is covariance propagation law (5-9), and the uncertainty synthesis does not need to apply the interpretation of A/B classification mechanically. Here is a simple example to illustrate this principle, which is also a comparison with the traditional practice.

For example, four points A, B, C and D are located on a straight line (Fig 7), and the observation data of distances obtained by geodimeter [14,15] are shown in Table 5. Please solve the final measured values of each line segment and the uncertainty of each error.

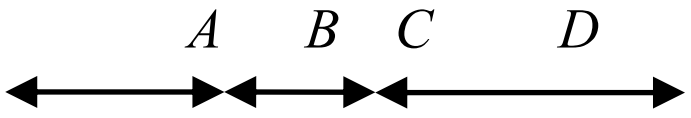

**Figure 7. Distances measurement**

Table 5: Observed values

| | Line segment | Observed values |
|---|---|---|
| **1** | AB | $x_1 = 39.8538m$ |
| **2** | BC | $x_2 = 159.957m$ |
| **3** | CD | $x_3 = 320.0015m$ |
| **4** | AC | $x_4 = 199.8117m$ |
| **5** | BD | $x_5 = 479.9601m$ |
| **6** | AD | $x_6 = 519.8149m$ |

Using $y_1$, $y_2$ and $y_3$ to express the final measured values of AB, BC and CD respectively, and using $k$ to express the measured value of the additive constant error of geodimeter, the observation error equation is

$$\begin{pmatrix} v_1 \\ v_2 \\ v_3 \\ v_4 \\ v_5 \\ v_6 \end{pmatrix} = \begin{pmatrix} x_1 \\ x_2 \\ x_3 \\ x_4 \\ x_5 \\ x_6 \end{pmatrix} - \begin{pmatrix} 1 & 0 & 0 & 1 \\ 0 & 1 & 0 & 1 \\ 0 & 0 & 1 & 1 \\ 1 & 1 & 0 & 1 \\ 0 & 1 & 1 & 1 \\ 1 & 1 & 1 & 1 \end{pmatrix} \begin{pmatrix} y_1 \\ y_2 \\ y_3 \\ k \end{pmatrix} \tag{7-3}$$

According to the least square method, there are:

$$\begin{pmatrix} y_1 \\ y_2 \\ y_3 \\ k \end{pmatrix} = \left[ \begin{pmatrix} 1 & 0 & 0 & 1 & 0 & 1 \\ 0 & 1 & 0 & 1 & 1 & 1 \\ 0 & 0 & 1 & 0 & 1 & 1 \\ 1 & 1 & 1 & 1 & 1 & 1 \end{pmatrix} \begin{pmatrix} 1 & 0 & 0 & 1 & 0 & 1 \\ 0 & 1 & 0 & 1 & 1 & 1 \\ 0 & 0 & 1 & 0 & 1 & 1 \\ 1 & 1 & 1 & 1 & 1 & 1 \end{pmatrix}^{\mathrm{T}} \right]^{-1} \begin{pmatrix} 1 & 0 & 0 & 1 & 0 & 1 \\ 0 & 1 & 0 & 1 & 1 & 1 \\ 0 & 0 & 1 & 0 & 1 & 1 \\ 1 & 1 & 1 & 1 & 1 & 1 \end{pmatrix} \begin{pmatrix} x_1 \\ x_2 \\ \vdots \\ x_6 \end{pmatrix}$$

$$= \frac{1}{4} \times \begin{pmatrix} 1 & -2 & -1 & 1 & -1 & 2 \\ -2 & 1 & -2 & 1 & 1 & 1 \\ -1 & -2 & 1 & -1 & 1 & 2 \\ 2 & 2 & 2 & 0 & 0 & -2 \end{pmatrix} \begin{pmatrix} x_1 \\ x_2 \\ \vdots \\ x_6 \end{pmatrix} \tag{7-4}$$

Substituting the numerical values of all observed values into equation (7-4), we get:

$$\begin{pmatrix} y_1 \\ y_2 \\ y_3 \\ k \end{pmatrix} = \begin{pmatrix} 39.8549 \\ 159.9583 \\ 320.0030 \\ -0.0013 \end{pmatrix} (m) \tag{7-5}$$

For the traditional measurement theory, the next step is to substitute (7-5) into (7-3), and six residual $v_i$ are obtained. Then $\sigma(x)$ is obtained by Bessel formula $\sigma(x) = \sqrt{\frac{\sum_{i=1}^{n} v_i^2}{n-t}}$, and $\sigma(y_1)$, $\sigma(y_2)$, $\sigma(y_3)$ and $\sigma(k)$ are obtained by covariance propagation law. Finally, $\sigma(y_1)$、$\sigma(y_2)$ $\sigma(y_3)$ and $\sigma(k)$ are called as precision or uncertainty of Type A, but the uncertainty of Type B is almost impossible to discuss.

However, from the perspective of the new conceptual theory, there are three conceptual troubles in the above variance submission process: 1. The degree of the freedom $n-t$ is too small, so it is meaningless to apply Bessel formula. 2. In Table 5, each observed value $x_i$ is a numerical value, and according to equation (7-5), each measured value $y_j$ is also a numerical value, so, their variances should be 0. 3. The contribution of the covariance between the errors of each observation value $x_i$ has not been taken into account at all (uncertainty synthesis issue).

The following is the variance submission process of the new conceptual theory for this case.

Taking the total differential of equation (7-5), the error propagation equation is obtained as follows:

$$\begin{pmatrix} \Delta y_1 \\ \Delta y_2 \\ \Delta y_3 \\ \Delta k \end{pmatrix} = \frac{1}{4} \times \begin{pmatrix} 1 & -2 & -1 & 1 & -1 & 2 \\ -2 & 1 & -2 & 1 & 1 & 1 \\ -1 & -2 & 1 & -1 & 1 & 2 \\ 2 & 2 & 2 & 0 & 0 & -2 \end{pmatrix} \begin{pmatrix} \Delta x_1 \\ \Delta x_2 \\ \vdots \\ \Delta x_6 \end{pmatrix} \tag{7-6}$$

Applying covariance propagation law (5-9) to equation (7-6), the covariance propagation equation is obtained as follows:

$$\begin{pmatrix} \sigma^2_{\Delta y_1} & \sigma_{\Delta y_1 \Delta y_2} & \sigma_{\Delta y_1 \Delta y_3} & \sigma_{\Delta y_1 \Delta k} \\ \sigma_{\Delta y_2 \Delta y_1} & \sigma^2_{\Delta y_2} & \sigma_{\Delta y_2 \Delta y_3} & \sigma_{\Delta y_2 \Delta k} \\ \sigma_{\Delta y_3 \Delta y_1} & \sigma_{\Delta y_3 \Delta y_2} & \sigma^2_{\Delta y_3} & \sigma_{\Delta y_3 \Delta k} \\ \sigma_{\Delta k \Delta y_1} & \sigma_{\Delta k \Delta y_2} & \sigma_{\Delta k \Delta y_3} & \sigma^2_{\Delta k} \end{pmatrix} = \frac{1}{16} \times \begin{pmatrix} 1 & -2 & -1 & 1 & -1 & 2 \\ -2 & 1 & -2 & 1 & 1 & 1 \\ -1 & -2 & 1 & -1 & 1 & 2 \\ 2 & 2 & 2 & 0 & 0 & -2 \end{pmatrix} \mathbf{D}(\Delta \mathbf{X}) \begin{pmatrix} 1 & -2 & -1 & 2 \\ -2 & 1 & -2 & 2 \\ -1 & -2 & 1 & 2 \\ 1 & 1 & -1 & 0 \\ -1 & 1 & 1 & 0 \\ 2 & 1 & 2 & -2 \end{pmatrix} \tag{7-7}$$

The acquisition process of the covariance matrix $\mathbf{D}(\Delta\mathbf{X})$ is as follows.

For the observed value $x_i$, its error $\Delta x_i$ is composed of three parts: additive constant error $K$, multiplication constant error $R$ and uneven indexing error $c_i$. That is:

$$\Delta x_i = K + R \cdot x_i + c_i \tag{7-8}$$

Its variance is

$$\sigma^2_{\Delta xi} = \sigma^2_K + x_i^2 \cdot \sigma^2_R + \sigma^2_c \tag{7-9}$$

The $\sigma_K$, $\sigma_R$ and $\sigma_c$ are obtained by consulting instrument instructions or the tolerance standard in instrument specification. Furthermore, according to formula (5-2), $\mathbf{D}(\Delta\mathbf{X})$ can be deduced as follows:

$$\mathbf{D}(\Delta\mathbf{X}) = E \begin{pmatrix} \Delta x_1 \\ \Delta x_2 \\ \vdots \\ \Delta x_6 \end{pmatrix} \begin{pmatrix} \Delta x_1 & \Delta x_2 & \cdots & \Delta x_6 \end{pmatrix}$$

$$= \begin{pmatrix} \sigma^2_K + x_1^2 \cdot \sigma^2_R + \sigma^2_c & \sigma^2_K + x_2 x_1 \cdot \sigma^2_R & \cdots & \sigma^2_K + x_6 x_1 \cdot \sigma^2_R \\ \sigma^2_K + x_1 x_2 \cdot \sigma^2_R & \sigma^2_K + x_2^2 \cdot \sigma^2_R + \sigma^2_c & \cdots & \sigma^2_K + x_6 x_2 \cdot \sigma^2_R \\ \vdots & \vdots & \vdots & \vdots \\ \sigma^2_K + x_1 x_6 \cdot \sigma^2_R & \sigma^2_K + x_2 x_6 \cdot \sigma^2_R & \cdots & \sigma^2_K + x_6^2 \cdot \sigma^2_R + \sigma^2_c \end{pmatrix} \tag{7-10}$$

Finally, the covariance matrix $\mathbf{D}(\Delta\mathbf{Y})$ is obtained by substituting the equation (7-10) into equation (7-7), where $\sigma(\Delta y_j)$ is called as uncertainty.

Assuming that there are $\sigma_K = \pm 2mm$, $\sigma_R = \pm 1 \times 10^{-6}$ and $\sigma_c = \pm 1mm$, we can get:

$$\begin{pmatrix} \sigma^2_{\Delta y_1} & \sigma_{\Delta y_1 \Delta y_2} & \sigma_{\Delta y_1 \Delta y_3} & \sigma_{\Delta y_1 \Delta k} \\ \sigma_{\Delta y_2 \Delta y_1} & \sigma^2_{\Delta y_2} & \sigma_{\Delta y_2 \Delta y_3} & \sigma_{\Delta y_2 \Delta k} \\ \sigma_{\Delta y_3 \Delta y_1} & \sigma_{\Delta y_3 \Delta y_2} & \sigma^2_{\Delta y_3} & \sigma_{\Delta y_3 \Delta k} \\ \sigma_{\Delta k \Delta y_1} & \sigma_{\Delta k \Delta y_2} & \sigma_{\Delta k \Delta y_3} & \sigma^2_{\Delta k} \end{pmatrix} = \begin{pmatrix} 0.75 & 0.01 & 0.26 & -0.50 \\ 0.01 & 0.78 & 0.05 & -0.50 \\ 0.26 & 0.05 & 0.85 & -0.50 \\ -0.50 & -0.50 & -0.50 & 5.00 \end{pmatrix} (mm^2)$$

Therefore, the uncertainties are:

$$\sigma_{\Delta y_1} \approx \pm 0.9mm \quad \sigma_{\Delta y_2} \approx \pm 0.9mm \quad \sigma_{\Delta y_3} \approx \pm 0.9mm \quad \sigma_{\Delta k} \approx \pm 2.2mm$$

It can be seen that there is $\sigma_{\Delta y_j} < \sigma_{\Delta x_i}$. Besides, it can be seen that the indication errors$\Delta x_i$ not only lead to the dispersion of repeated observations $x_1 \sim x_6$, but also lead to their overall deviation. If we entangle in the influence characteristics (A/B classification) of indication error on repeated observations, it will not only be unable to express, but also will not help to solve the problem. Moreover, whether errors $K, Rx_i, c_i, \Delta x_i$ or error $\Delta y_j$ are all deviations, and have variances used to evaluate their probability intervals, while the systematic error without variance does not exist.

## 8. Conclusions

In short, different from the traditional theory, the new theory follows rigorous mathematical concepts and regards both the observed value and measured value as constants, and the error and true value as random variables, so that the conceptual logic of new theory has changed in an overall way (shown in Table 6). First, the variance (standard deviation) or the uncertainty is the evaluation of probability interval of a single error (deviation), but not the dispersion of a measured value, and any regular error's size degree can be evaluated by them. Second, any error follows a random distribution, has variance which can be used to evaluate its size, and cannot be classified as the systematic error or the random error. Third, the error synthesis follows the algebraic rule, the variance synthesis follows the probability principle, and there is no need to use those old concepts such as systematic error, random error, precision, trueness and accuracy. As a result, the revision of measurement textbooks and metrological terminology will become issues that need to be re-discussed in the future.

Table 6: Conceptual logic difference between the two theories.

| Traditional measurement theory | New conceptual measurement theory |
| --- | --- |
| The measured value is a random variable because it changes randomly in repeated measurements. | The measured value is a numeric value, and is a constant. |
| The true value is a constant because it remains constant during repeated measurements. | The true value is unknown, and is a random variable that needs to be described by a probability range. |
| The best measured value is given by random error analysis. | The best measured value is given by analyzing the randomness of errors. |
| With the true value as the reference center, submit the evaluation of the deviation and dispersion (reproducibility) of the measured value. | With the measured value as the reference center, submit the evaluation value of the probability range that the true value deviates from the measured value. |
| The errors are divided into systematic and random classifications. | The errors are not divided into systematic and random classifications. |
| …… | …… |

## References

[1] JCGM 100:2008, Guide to the Expression of Uncertainty in Measurement (GUM).
[2] JJF1059-2012, Evaluation and Expression of Uncertainty in Measurement.
[3] JCGM 200:2012, International vocabulary of metrology — Basic and general concepts and associated terms (VIM).
[4] JJF1001-2011, General Terms in Metrology and Their Definitions.
[5] GB/T14911-2008, Basic Terms of Surveying and Mapping.
[6] Huang Hening. A unified theory of measurement errors and uncertainties. Measurement Science and Technology, 2018, 29(12).
[7] Xiao-ming Ye, Mo Ling, Qiang Zhou, Wei-nong Wang, Xue-bin Xiao. "The New Philosophical View about Measurement Error Theory". Acta Metrologica Sinica, vol. 36, pp. 666-670, 2015.
[8] Xiao-ming Ye , Xue-bin Xiao, Jun-bo Shi, Mo Ling. "The new concepts of measurement error theory", Measurement, vol.83, pp. 96-105, 2016.
[9] Xiao-ming Ye, Hai-bo Liu, Mo Ling, Xue-bin Xiao. "The new concepts of measurement error's regularities and effect characteristics", Measurement, vol.126, pp.65-71, 2018.
[10] Xiao-ming Ye, Shi-jun Ding. "Comparison of variance concepts interpreted by two measurement theories". Journal of Nonlinear and Convex Analysis, vol.20, pp.1307-1316, 2019.
[11] Yetai Fei. Error Theory and Data Processing. Machinery Industry Press, 2016 3.
[12] Depei Ye. Understanding, Evaluation and Application of Measurement Uncertainty. China Metrology Publishing House, 2007 10.
[13] School of Geodesy and Geomatics, Wuhan University. Error Theory and Foundation of Surveying Adjustment[M]. Wuhan University Press, 2016 6.
[14] JJG703-2003, Electro-optical Distance Meter (EDM instruments).
[15] ISO 17123-4:2012,Optics and optical instruments -- Field procedures for testing geodetic and surveying instruments -- Part 4: Electro-optical distance meters (EDM measurements to reflectors).